# Observation of a Room Temperature Two-dimensional Ferroelectric Metal

Mao Ye[1†], Songbai Hu[1†], Shanming Ke[2*], Yuanmin Zhu[3,4], Yubo Zhang[1], Lin Xie[1], Yuan Zhang[5], Dongwen Zhang[6], Zhenlin Luo[7], Meng Gu[4,3], Jiaqing He[1], Peihong Zhang[1,8], Wenqing Zhang[1], Lang Chen[1,3*]

[1] Department of Physics, Southern University of Science and Technology, Shenzhen 518055, China
[2] School of Materials Science and Engineering, Nanchang University, Nanchang 330031, China
[3] Academy for Advanced Interdisciplinary Studies, Southern University of Science and Technology, Shenzhen 518055, China
[4] Department of Materials Science and Engineering, Southern University of Science and Technology, Shenzhen 518055, China
[5] School of Materials Science and Engineering, Xiangtan University, Hunan 411105, China
[6] College of Science, National University of Defense Technology, Hunan 410073, China
[7] National Synchrotron Radiation Laboratory, University of Science and Technology of China, Hefei 230026, China
[8] Department of Physics, University at Buffalo, State University of New York, Buffalo, USA
*Correspondence to: chenlang@sustech.edu.cn and ksm@ncu.edu.cn
† These authors contributed equally

**Materials with reduced dimensions have been shown to host a wide variety of exotic properties and novel quantum states that often defy textbook wisdom[1-5]. Ferroelectric polarization and metallicity are well-known examples of mutually exclusive properties that cannot coexist in bulk solids because the net electric field in a metal can be fully screened by free electrons[6]. An atomically thin metallic layer capped by insulating layers has shown decent conductivity at room temperature[7]. Moreover, a penetrating polarization field can be employed to induce an ion displacement and create an intrinsic polarization in the metallic layer. Here we demonstrate that a ferroelectric metal can be artificially synthesized through imposing a strong polarization field in the form of ferroelectric/unit-cell-thin metal superlattices. In this way the symmetry of an atomically thin conductive layer can be broken and manipulated by a neighboring polar field, thereby forming a two-dimensional (2D) ferroelectric metal. The fabricated of $(SrRuO_3)_1/(BaTiO_3)_{10}$ superlattices exhibit ferroelectric polarization in an atomically thin layer with metallic conductivity at room temperature. A multipronged investigation combining structural analyses, electrical measurements, and first-principles electronic structure calculations unravels the coexistence of 2D electrical conductivity in the $SrRuO_3$ monolayer accompanied with the electric polarization. Such 2D ferroelectric metal paves a novel way to engineer a quantum multi-state with unusual coexisting properties, such as ferroelectrics, ferromagnetics and metals, manipulated by external fields[8,9].**

The coexistence of two/multiple mutually exclusive states in low-dimensional materials is attracting considerable attention owing to their novel states of matter and fascinating phenomena undiscovered in bulk[1-4,10-14]. Ferroelectric metal possessing both ferroelectricity and metallicity, as an incredible paradigm of multiferroics, is firstly introduced by Anderson and Blount over five decades ago[15]. Since then, only a few polar metals have been identified[6,16-20]. These non-centrosymmetric structures with metallicity are still not technically regarded as actual ferroelectrics, because the macroscopic polarization cannot be switched by an external electric field. Polarization field has proven to be a powerful strategy in acquiring emergent properties such as 2D electron or hole gas, polar vortices, strongly anisotropic polarization-induced conductivity[3,4,11-14,21-23]. However, there are few examples on the accomplishment of ferroelectric metal. In this report, we demonstrate that ferroelectric metal can be artificially synthesized through imposing a polarization field in the form of ferroelectric/unit-cell-thin metal superlattices. In this way the symmetry of an atomically thin conductive layer can be broken and manipulated by a neighboring polar field, thereby forming a 2D ferroelectric metal.

We chose the $(SrRuO_3)_1/(BaTiO_3)_{10}$ superlattices made of a periodically arranged monolayer $SrRuO_3$ and ten-unit cells of traditional ferroelectric $BaTiO_3$. It was believed that $SrRuO_3$ becomes insulating down to three-unit cells[24,25]. Recent studies show one-unit cell $SrRuO_3$ can be conductive by capping $SrTiO_3$ (Extended Data Fig. 1)[7]. The ferroelectric $BaTiO_3$ layer provides a polarization field penetrating the atomically thin $SrRuO_3$ layer, which will induce a Ru-O ion displacement, and modify the transport behavior. We present both quantitative experimental measurements and first-principles calculations to demonstrate the coexistence of conductivity and polarization in $SrRuO_3$. Indeed, we find that the 2D electrical conductivity is confined in the atomically thin $SrRuO_3$, meanwhile, the inversion symmetry is broken due to the presence of ferroelectric $BaTiO_3$.

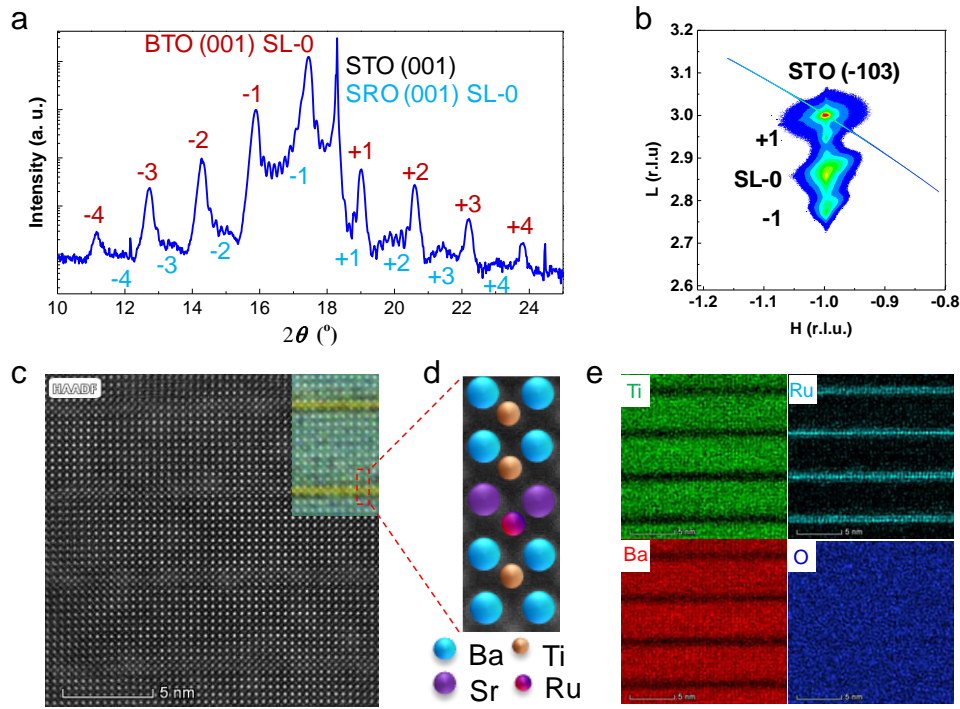

**Figure 1. Structural characterization of the $(SrRuO_3)_1/(BaTiO_3)_{10}$ superlattices. a**, XRD $\theta$-$2\theta$ scan. **b**, reciprocal space mapping (RSM) around (-103) peak. **c**, High-angle annular dark-field (HAADF) STEM images (the inset shows the atomic EDS mapping of two neighbor $SrRuO_3$ layers). **d**, The magnified image from the area marked with a red dashed rectangle in **c**. **e**,

the Ti, Ru, Ba and O elemental mapping of the $(SrRuO_3)_1/(BaTiO_3)_{10}$ superlattices, confirming a coherent growth of the superlattices.

We grew the $(SrRuO_3)_1/(BaTiO_3)_{10}$ superlattices using pulsed laser deposition with *in-situ* reflection high energy electron diffraction (RHEED) on $TiO_2$-terminated $SrTiO_3$ (001) substrates (Extended Data Fig. 2). The surface topography of the superlattices follows the terraces of the substrate, indicating a step-flow growth of the thin films (Extended Data Fig. 3). The structure of the $(SrRuO_3)_1/(BaTiO_3)_{10}$ superlattices is characterized by both synchrotron XRD and high-resolution scanning transmission electron microscopy (STEM) (Fig. 1). Figure 1a shows at least four orders of satellite peaks on the two sides of the SL-0 (SL stands for superlattices) peak were observed. Those small shoulders next to the $BaTiO_3$ peaks are identified as $SrRuO_3$ due to its smaller *c*-lattice constant than $BaTiO_3$. The periodicity is calculated to be 44.93 Å using the distance of the satellite peaks according to the Bragg's law, which well matches the designed thickness (44.31 Å without strain). The measured *c*-lattice constant of $BaTiO_3$ layer (4.09 Å) is larger than the bulk value (4.04 Å), indicating that the superlattices are compressively strained by the $SrTiO_3$ substrate. The RSM (Fig. 1b) proves that the superlattices are fully-strained as no horizontal shift is detected along the (-103) orientation. Thus, the $BaTiO_3$ layer is elongated along the out-of-plane direction and exhibits a higher *c*/*a* ratio than the bulk. The atomically sharp layer $SrRuO_3$ confined by $BaTiO_3$ is imaged directly using the double aberration-corrected TEM. The lower-magnification HAADF-STEM image exhibits ten repetitions of the building blocks (Extended data Fig. 4). The $(SrRuO_3)_1/(BaTiO_3)_{10}$ superlattices structure is captured distinctly by scanning transition electron microscopy with high-angle annular-dark-field detectors (HAADF-STEM) where no defects or stacking fault is observed (Fig. 1c). The inset displays an atomic EDS map overlapped on the same area, in which the atomically sharp $SrRuO_3$ monolayer was observed distinctively. Base on that, Figure 1d builds up the atom's arrangement across the $(SrRuO_3)_1/(BaTiO_3)_{10}$ heterostructure. The respective EDS elemental mapping shows no inter-diffusion of Ti, Ru, and Ba between the $SrRuO_3$ and the $BaTiO_3$ layer (Fig. 1e). In that sense, we have obtained fully-strained $(SrRuO_3)_1/(BaTiO_3)_{10}$ superlattices with an atomically sharp interface on $SrTiO_3$ substrate.

The ferroelectricity of the $(SrRuO_3)_1/(BaTiO_3)_{10}$ superlattices was investigated by piezoresponse force microscopy (PFM), second harmonic generation (SHG) and STEM separately (Fig. 2). The local out-of-plane piezoresponse of the superlattices was measured on the bare surface as a function of a voltage applied to a conductive tip (Fig. 2a). That amplitude hysteresis loop and butterfly phase curves show clearly the symmetric ferroelectric switching behavior of the superlattices. The switching voltages were approximately 2 V. Figures 2b & c show the PFM poling map written at +5 V on the larger box and −5 V on the smaller box. A 180° phase contrast between the square domain patterns qualitatively demonstrates the appearance of the up- and down-polarization, respectively, which illustrates that a switchable polarization can exist in the $(SrRuO_3)_1/(BaTiO_3)_{10}$ superlattices at room temperature. The unpoled region beyond the written box displays a spontaneous up-polarization, which can be attributed to the fact that the $SrRuO_3$ prefers to end up with SrO-termination when grown on $TiO_2$-terminated $SrTiO_3$[26,27]. In addition, we investigated the presence of polar displacements through crystal structure asymmetries, i.e. optical second harmonic generation (SHG) measurement, to establish further the ferroelectricity. The results are shown in Fig. 2d. The SHG signals indicate clearly a polar structure in the $(SrRuO_3)_1/(BaTiO_3)_{10}$ superlattices. Both the *p*- and *s*-polarized SHG signals are fitted by detailed theoretical modeling (see Methods). The well-fitted results show that the superlattices possess a net *mm2* point group with an out-of-plane polarization[28].

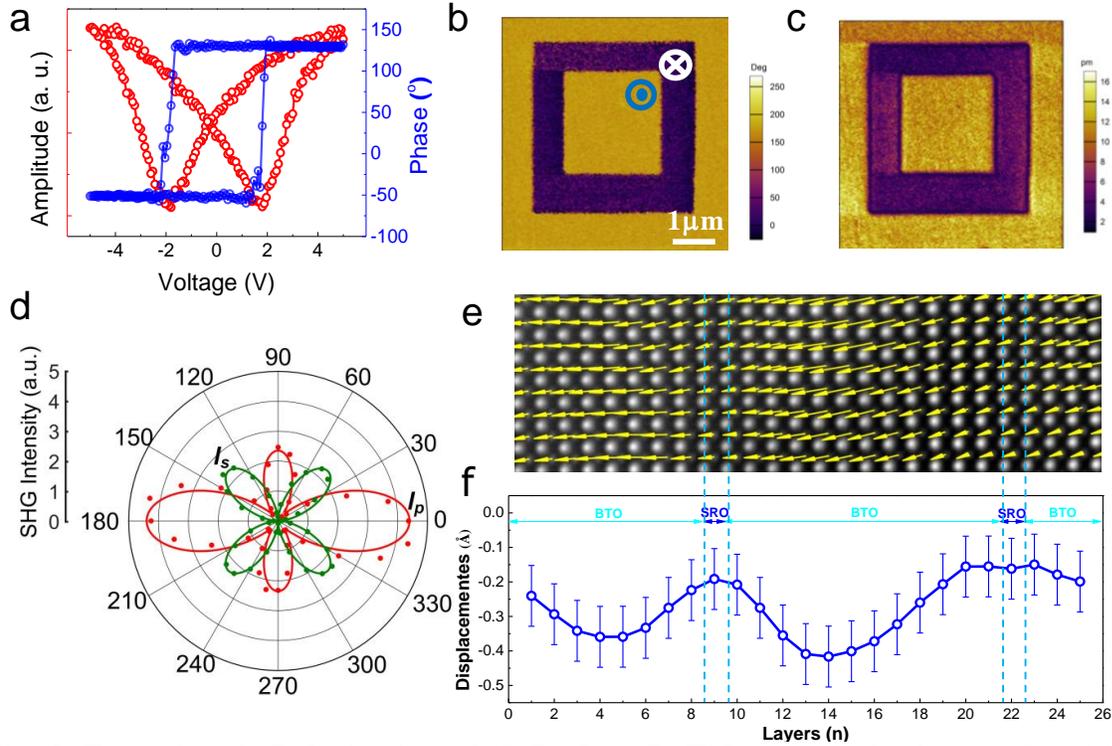

**Fig. 2. Ferroelectric Polarization of $(SrRuO_3)_1/(BaTiO_3)_{10}$ superlattices. a**, Local hysteretic behavior of the $(SrRuO_3)_1/(BaTiO_3)_{10}$ superlattices characterized by PFM phase and amplitude scanning on the bare surface as a function of the bias. **b**, **c**, Phase and Amplitude of PFM poling map written at +/- 5 V. **d**, Polar plots of SHG intensity vs. incident polarization angle for the $(SrRuO_3)_1/(BaTiO_3)_{10}$ superlattices. The red and blue circular symbols are the measured *p*-, and *s*-polarized SHG signals, respectively. The solid lines are the fitting results for them. **e**, Filtered Cross-sectional HR-STEM image with an overlay of the polar displacement vectors (indicated by yellow arrows), fitted with Gaussian peaks for distance calculation in the $(SrRuO_3)_1/(BaTiO_3)_{10}$ superlattices. **f**, The polar displacement derived from (**e**) along the out-of-plane direction. The error bar shows the standard deviations of the averaged displacement for each atomic layer.

Furthermore, we investigated polar displacements at the atomic scale by observing the centrosymmetric breaking of $TiO_6$ and $RuO_6$ octahedra. To address this, aberration-corrected HAADF-STEM imaging was carried out, which allows to directly quantify the precise atom positions in the $(SrRuO_3)_1/(BaTiO_3)_{10}$ superlattices. As shown in Fig. 2e and Extended data Fig. 5, significant Ti deviation, i.e., relative shift along the out-of-plane direction between Ti and the mass center of a rectangle formed by its four Ba neighbors was observed. Significantly, the Ru deviation was captured in atomically thin $SrRuO_3$, which is direct evidence for the polarization in $SrRuO_3$. This ion displacement arises from the penetrated polarization field from the $BaTiO_3$ layers[18]. The superlattices show an up-polarization in the as-grown state, which is consistent with the PFM measurements (Fig. 2b). Figure 2f shows the cation displacement profile of $TiO_6$ and $RuO_6$ octahedra layer by layer across the interfaces. The Ti deviations are found to be as large as 0.4 Å. The Ru ion displacements in SrRuO$_3$ layers has a significant ~ 0.2 Å which amounts to 5% of the *c*-lattice parameter (~ 4 Å for bulk SrRuO$_3$). These TEM results are in agreement with those XRD, SHG and PFM data and further verify the presence of the polarization in the atomically thin SrRuO$_3$ layer.

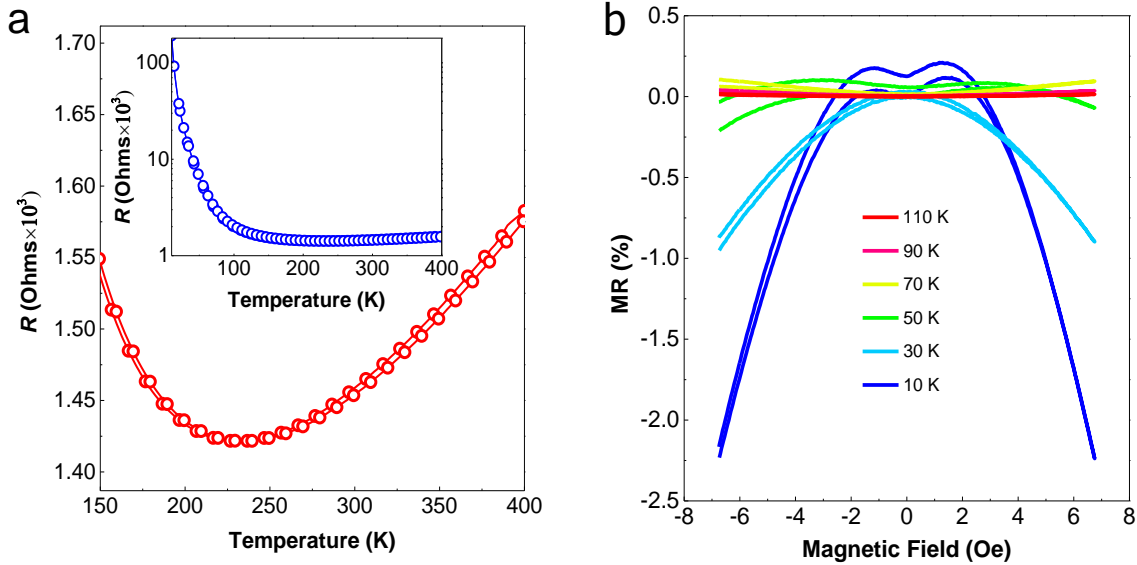

**Figure 3. Electrical transport measurements of the (SrRuO$_3$)$_1$/(BaTiO$_3$)$_{10}$ superlattices. a**, The temperature-dependent resistance of the (SrRuO$_3$)$_1$/(BaTiO$_3$)$_{10}$ superlattices demonstrating a metallic behavior in the temperature range from 225 to 400 K, however, below the temperature 225 K, illustrating an insulating characteristic. **b**, MR curves at different temperatures. Negative MR appeared below 70 K, while above that positive MR was observed.

In our superlattices, SrRuO$_3$ monolayer clamped by BaTiO$_3$ ferroelectric layers were found to be metallic at room temperature. We carry out macroscopic electronic transport measurement by a standard four-terminal method over the temperature range from 10 K to 400 K. The wire electrodes were made to penetrate the interface for collecting the bulk rather than the surface current. The results are shown in Fig. 3a. The resistance increases with a temperature between 225 K and 400 K, exhibiting a metallic behavior. We find it is consistent with ref. [29] that the electrons are confined in the monolayer SrRuO$_3$. However, it decreases with the temperature below 225 K, illustrating an insulating characteristic. Such a metal-to-insulator transition is probably attributed to the localization of the charge carriers[7]. A similar phenomenon is observed in (SrRuO$_3$)$_1$/(SrTiO$_3$)$_6$ superlattices (Extended Data Fig. 1). To further study the transport behavior, the magnetoresistance (MR) of the superlattices was measured at different temperatures. The negative MRs at low temperature probably arise from the broken time-reversal symmetry at the presence of an external magnetic field, which leads to the destruction of weak localization of electrons[30]. Such behavior was considered to be indirect evidence of ferromagnetism (for more magnetization measurement please see Extended Data Fig. 6). Moreover, the absolute value of negative MR decreases with the increase of the temperature, consistent with the characteristics of weak localization systems[30]. However, a small positive MR is also observed above 70 K, which likely originates from the spin splitting of conduction electron energies[29].

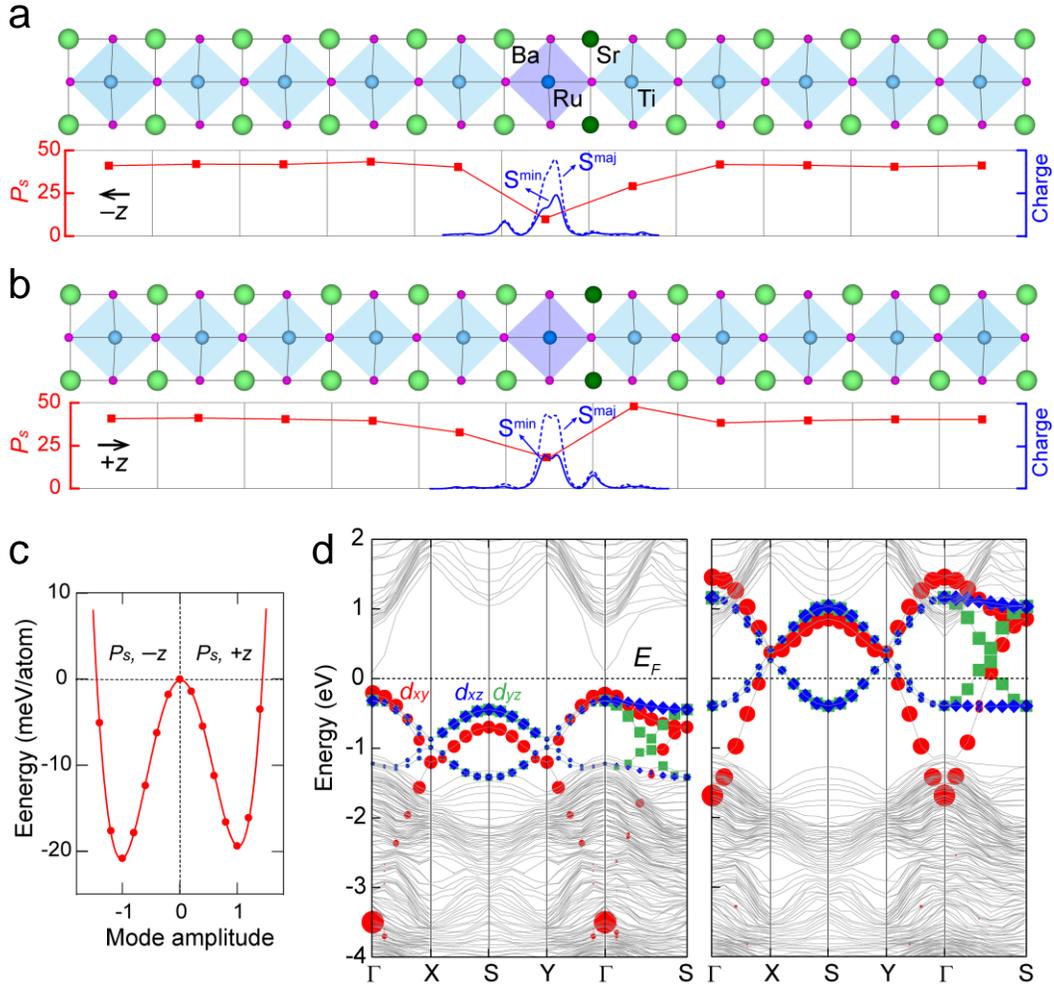

**Figure 4. Ferroelectric polarization and metallic conductivity predicted by density functional theory. a**, The upper panel is the superlattices model for polarization along the $-z$ direction. The lower panel shows the layer-resolved polarization magnitudes (in $\mu C/cm^2$) and charge density. The integrated planar charge is for electrons from the Fermi level to $-1$ eV below, for both the majority- and minority-spin channels of the ferromagnetic phase. **b**, The same as in subplot (**a**), except for the polarization is along the $+z$ direction. **c**, Total energy as a function of the polar distortion amplitude. The double-well structure is asymmetric for $+z$ and $-z$ distortions. **d**, Energy dispersion relations within the $RuO_2$ planar directions for the half-metallic state. The left and right panels are for the majority- and minority-spin, respectively. Wavefunctions of five Ru-$4d$ orbitals are superimposed as colored symbols.

The microscopic mechanism of ferroelectric polarization and metallic conductivity of the $(SrRuO_3)_1/(BaTiO_3)_{10}$ superlattices is simulated by the density functional theory, as shown in Fig. 4. The out-of-plane polarization persists in the central $BaO-RuO_2-SrO$ layer, although it is gradually reduced in general (Figs. 4a & b). Microscopically, the polarization persistence is indirectly driven by the polar atomic displacements of the adjoined $BaTiO_3$ layers (also see Extended Data Fig. 7 and corresponding discussions for details). The polarization bi-stabilities along the upward and downward directions are clearly asymmetric due to $BaO-RuO_2$ and $SrO$-

RuO$_2$ symmetry breaking across the interface (Fig. 4c). Metallicity is derived from the strongly confined 2D electric conductivity within the RuO$_2$ plane (Figs. 4a & b). Specifically, the metallic conductivity is mainly contributed by the electrons in the minority-spin channel which are derived from the partially filled Ru-4$d$ orbitals (i.e., one electron occupies three t$_{2g}$ orbitals) and of bonding state, although the minority electrons are much less than the majority electrons, which are from the fully occupied Ru-4$d$ orbitals (i.e., three electrons in three t$_{2g}$ orbitals) and of anti-bonding state (Fig. 4d). Our theoretical predictions unambiguously confirm the 2D ferroelectric metal in atomically thin SrRuO$_3$ in agreement with the experimental measurements (Fig. 2 and Fig. 3).

To summarize, we introduced strong ferroelectric BaTiO$_3$ layers to polarize a one-unit-cell thin metallic SrRuO$_3$ layer and achieved successfully a novel ferroelectric metal in (SrRuO$_3$)$_1$/(BaTiO$_3$)$_{10}$ superlattices. According to our experimental and theoretical results, the conductivity originates solely from the atomically thin SrRuO$_3$. The polar displacements exist not only in BaTiO$_3$, but also in the atomically thin SrRuO$_3$, demonstrating a 2D ferroelectric metal. The approach not only creates a new platform for the 2D polar metal system, but also shows an emerging route to discover novel multiferroics with the coexistence of incompatible physical properties, e. g., ferroelectrics, ferromagnetic and metals. The proposed structure presents a tantalizing opportunity to integrate the competing order parameters such as charge, orbital, spin and lattice, and paves an avenue to multifunctional oxide devices, which can be used for future memories and nanoelectronics.

**Acknowledgments** We acknowledge support by the National Natural Science Foundation of China (11604140, and 11804145), the Science and Technology Research Items of Shenzhen (JCYJ20170412153325679, JCYJ20180504165650580, and JCYJ20170817110302672), the Natural Science Foundation of Guangdong Province of China (2018A030310221) and the

Introduced Innovative R&D Team of Guangdong (2017ZT07C062). S.M.K. would like to thank the financial support from Nanchang University. L.C. and W.Q.Z. acknowledge the support from Shenzhen Pengcheng-Scholarship program. We acknowledge the Materials Characterization and Preparation Center of SUSTech for their help.

**Author contributions** The project was conceived by M.Y., S.M.K. and L.C. Films were fabricated by M.Y. Synchrotron XRD and RSM were carried out and analyzed by Z.L.L. and SBH. STEM was carried out by Y.M.Z. and L.X. under the guidance of M.G. and J.Q.H., and analyzed by Y.M.Z., L.X., S.B.H. and M.Y. AFM and PFM were carried out and analyzed by M.Y. and SBH. SHG was carried out and analyzed by Y.Z. and D.W.Z. Electronic transport properties were measured M.Y. and S.B.H. Theoretical calculations were carried out by Y.B.Z. with the guidance of P.H.Z. and W.Q.Z. M.Y., S.B.H., S.M.K. and C.L. wrote the manuscript, and all the authors participated discussions and writing.

**Competing Interests** The authors declare no conflict of interests.